\documentclass[12pt]{article}
\input epsf.sty

\newcommand{\sect}[1]{\setcounter{equation}{0}\section{#1}}

\newcommand{\eqn}[1]{(\ref{#1})}
\newcommand{\br}{\nonumber \\}
\newcommand{\be}{\begin{equation}}
\newcommand{\ee}{\end{equation}}
\newcommand{\bea}{\begin{eqnarray}}
\newcommand{\eea}{\end{eqnarray}}
\newcommand{\p}{\partial}

\newcommand{\plb}[3]{Phys. Lett. {\bf B#1} ({#2}) {#3}}
\newcommand{\prd}[3]{Phys. Rev. {\bf D#1} ({#2}) {#3}}
\newcommand{\hepth}[1]{{\tt hep-th/{#1}}}
\newcommand{\hepph}[1]{{\tt hep-ph/{#1}}}

\begin{document}
\renewcommand{\thefootnote}{\fnsymbol{footnote}}
\begin{titlepage}
\begin{flushright}
\end{flushright}
\vskip .7in
\begin{center}
{\Large \bf Spinflation from Geometric Tachyon} \vskip .7in
{\large Kamal L. Panigrahi $^a$\footnote{e-mail: {\tt
panigrahi@iitg.ernet.in}} and \large Harvendra Singh
$^b$\footnote{e-mail: \tt h.singh@saha.ac.in}} \vskip .2in
{$^a$}{\it Department of Physics\\
Indian Institute of Technology Guwahati,\\
Guwahati, 781 039, India} \vskip .2in
{$^b$}{\it Theory Division, Saha Institute of Nuclear Physics\\
1/AF Bidhannagar, Kolkata 700064, India} \vspace{.7in}
\begin{abstract}
\vskip .5in \noindent We study the assisted inflation scenario
from the rolling of $N$ BPS D3-brane into the NS5-branes, on a
transverse geometry of $R^3 \times S^1$, coupled to four
dimensional gravity. We assume that the branes are distributed
along $S^1$ and the probe D3-branes spin along $R^3$ plane.
Qualitatively this process is similar to that of N-tachyon
assisted inflation on unstable D-branes. We further study the
spinflation scenario numerically and analyze its effect.

\end{abstract}
\end{center}
\vfill

\end{titlepage}
\setcounter{footnote}{0}
\renewcommand{\thefootnote}{\arabic{footnote}}
\tableofcontents

\sect{Introduction} Inflation provides us a useful answer for the
homogeneity and isotropy of the present day observed universe
\cite{guth,linde}. It can solve dynamically the flatness and
horizon problem of the universe. Inflation should therefore be
emerged from any theory, e.g. string theory, considered to be a
fundamental theory. It is not surprising therefore that de Sitter
like inflationary vacua can be constructed in string theory
\cite{kklt}. Several attempts have been made for deriving
inflation from various string inspired models
\cite{kklmmt,dvali,mac}. There are in fact a large number of such
examples in string theory. One class of such models based on
open-string tachyon condensation  are capable of providing
slow-roll inflation, see
\cite{anup,sen3,Piao:2002vf,hs1,sinha,hs2,zamarias,anne,hsingh,panigrahi}.
The open-string tachyon is a unique candidate and has a
inflationary potential \cite{sen0,sen1}, but in general, the
simple type of models are plagued with the same large
$\eta$-problem as the conventional models and are not favoured for
slow-roll inflation, see \cite{kofman}. Although, this difficulty
can be avoided by allowing a large number of tachyons to
simultaneously roll down and {\it assist} the inflation
\cite{hs2,hsingh}. This naive idea is also known as assisted
inflation \cite{liddle}. Some recent developments on N-flation and
multi-brane inflation can be found in \cite{Dimo,
Ward:2007gs,Grimm:2007hs,Krause:2007jr,
Ahmad:2008eu,becker,becker1}. Other inflationary models based on
brane-intersection at special angles \cite{garcia}, the D3/D7
models where the distance modulus (like in our present study)
plays the roll of inflaton field \cite{keshav, chen} and some of
the race-track models driven by the K\"ahler modulus
\cite{racetreck}, are other interesting cases where the inflation
is omnipresent in string theory constructions, (see e.g.
\cite{Kehagias:1999vr,GomezReino:2002fs,Pajer:2008uy}). However,
one can see for a wider recent review \cite{McAllister:2007bg}.

A geometric tachyon is the geometrical interpretation of the
perturbative open string tachyon field in terms of the radial
distance between the in-falling D-brane into a stack of NS5-branes
\cite{kutasov1,Kutasov:2004ct}. Once this equivalence was
proposed, it was tested both from the view point of effective
field theory by looking at the Dirac-Born-Infeld action of the
probe D-brane in the NS5-brane geometry (and other D$p$-brane
background) \cite{Rbrane}, and from the full string theory view
point by looking at various boundary states of rolling branes in
this background \cite{NST-BS}. More recently in \cite{NS5}, it was
shown that one can get qualitatively all the behaviour of unstable
D-branes in flat space with that of BPS and non-BPS branes in
NS5-brane background when the NS5-branes have a transverse
geometry of $R^3\times S^1$. It was observed \cite{Kutasov:2004ct}
that the dynamics of D-branes, which are BPS in flat ten
dimensional spacetime, propagating in the background of $k$
NS5-branes on the transverse space $R^3\times S^1$ are remarkably
similar to that of BPS and non-BPS branes in ten dimensions.
Further, it was found out that what looks to be a BPS or non-BPS
branes in six dimensions is actually the same object-the BPS
D-brane in ten dimensions wrapped or unwrapped around the extra
$S^1$.

The cosmological solution has also been found out by looking at
the D3-brane motion into the NS5-brane geometry\cite{NS-cosm}. The
inflationary scenario has been worked out in this picture. In
particular the assisted inflation from the geometric tachyon has
been discussed in \cite{panigrahi}, where a large number of
D3-branes have been shown to roll simultaneously to get the slow
roll parameters. However it has been found out that the radion
should acquire trans-stringy vev. This also assumes the number of
D3-branes to be very large. We try to improve upon this situation
in the present paper by assuming that the D3-brane rolls and spins
simultaneously into the NS5-branes on $R^3\times S^1$. We look at
the assisted inflationary scenario in this background geometry by
allowing $N$ D3-branes to simultaneously roll and spin into this
geometry. Due to the extra spin on the D3-brane, the process of
`falling' into the NS5-branes will be delayed further, and there
is a fair chance of getting the slow roll parameters better. We
have examined this process of spinflation in the present paper.
Originally the idea of spin-flation in the brane-inflation  has
been studied in \cite{easson}.
 We shall be assuming that
both the D3-branes and NS5-branes are distributed along $S^1$ in
addition
to the other longitudinal directions and the rolling branes also
spin along the $R^3$-plane. We will show that even without the
spin along $R^3$ plane and for a small radius of the compact
circle along which the $N$ D3-branes are distributed, (and for a
very weak string coupling), we get the number of D3-branes
required for slow roll to be small indeed. On the other hand,
one has to keep the number of NS5-branes under control.
Nevertheless the vev of the geometric tachyon still seem to be
trans-stringy ($\langle\Phi\rangle > M_s$) which is not so bad as
that can be achieved by placing D-branes  far away from the
NS-branes. Then we assume that the D3-branes are spinning into the
NS5-branes, and study the effect of slow roll numerically. We will
show that at the initial stages the tachyons roll very slowly, and
afterwards the motion resembles with that of the zero-angular
momentum case.

The rest of the paper is organized as follows. In section-2, we
study the probe D3-brane motion into the NS5-brane on a transverse
space which is $R^3\times S^1$, where the branes are distributed
along $S^1$ of radius $R'>l_s$. For the range $z\gg R'$, where $z$
is the radial direction in $R^3$, we identify the radial mode, which
behaves
like the open string tachyon and write down the relevant potential
at a distance far away from the NS5-branes. Next we couple this
system with that of four dimensional gravity and solve the
equations of motion and find out the effective potential for
studying the slow roll. We write down the conditions for slow roll
and find out the number of D3-branes that we need for the assisted
inflation. Section-3 is devoted to study spinflation from the
geometric tachyon, where we assume a conserved angular momentum on
the D-brane along $R^3$. We write down the equations of motion and
study the spinflation numerically and compare the result with that
of zero angular momentum case. Finally in section-4 we present our
conclusions.

\sect{NS5-D3  brane system on $R^3\times S^1$}
\subsection{Geometric tachyon}
We start with $k$ NS5-branes on a transverse $R^3\times S^1$
space, which we will label by the coordinates $(\vec z,y)$, with
$\vec z\in R^3$ and $y\sim y+2\pi R'$ ($R'$ being the radius of
the $S^1$). The five-branes are located at the point $\vec z=y=0$.
The background geometry generated by this is
\cite{Callan:1991at,Gauntlett:1992nn}\bea ds^2 &=& dx_\mu
dx^\mu+h(\vec z,y)\left(d\vec z^2+dy^2\right)~,\cr
e^{2(\phi-\phi_0)} &=& h(\vec z,y)~,\cr {\cal
H}_{mnp}&=&-\epsilon_{mnpq}\partial^q \phi ~. \eea The $x^\mu\in
R^{5,1}$ label the worldvolume of the five-branes. $\phi_0$ is
related to the string coupling far from the five-branes,
$g_s=\exp\phi_0$. The harmonic function $h$ has the form \bea h =
1 + {k\alpha'\over 2R'|\vec z|}{\sinh(|\vec z|/R')\over\cosh(|\vec
z|/R')-\cos(y/R')}~.\eea For our purpose, we are interested in the
geometry very far away from the NS5-branes. So for $|z| \gg R'$,
the harmonic function is given by \bea h= 1+ k \alpha'/(2R' |z|)\
,\eea and thus the $y$-direction acts as an isometry direction.
The effective action on the world volume of D$p$-brane in this
background is governed
by the DBI action:
\begin{eqnarray}
S_{p} &=& -T_{p} \int d^{p+1} \xi~ e^{-(\phi
-\phi_0)}\sqrt{-\det(G_{ab} + B_{ab})} \cr & \cr &=& -\tau_{p}\int
d^{p + 1}x \frac{1}{\sqrt{h(Z)}} \sqrt{1 + h \p_a Z \p^a Z}
\end{eqnarray}
where in the first line $G_{ab}$ and $B_{ab}$ are the induced
metric and the $B$-field, respectively, onto the world volume of the
D$p$-brane. Comparing it with the open string tachyon effective
action
\bea {\mathcal S}_{tach} = - \int d^{p + 1}x ~ V(T) \sqrt{1
+ \p_{a} T \p^a T}, \label{act-tach} \eea
one gets the so called
radion-tachyon map: \bea dT/dZ  = \sqrt{h (Z)}= \sqrt{1+ Q^2
/Z} \eea where $Z$ is the radion field, which is tachyonic.
Solving the above differential equation, one gets \bea T = Z
\sqrt{1+ Q^2 /Z} + Q^2 \log(Z^{1/2} +(Q^2+Z)^{1/2})\eea
where $Q^2\equiv k \alpha'/(2R')$. This for large $Z$ gives
$T\simeq Z$. In this asymptotic region the tachyon potential becomes
\bea
V(T) ={1\over\sqrt{h(Z)}}\simeq 1-{Q^2\over 2 T} \ . \eea
\begin{figure}[!ht]
\leavevmode
\begin{center}
\epsfysize=10cm
\epsfbox{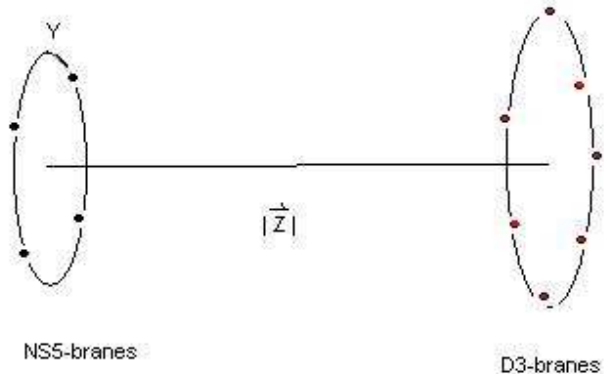}
\end{center}
\caption{\it $k$ NS5-branes and $N$ D3-branes are uniformly spread
over transverse $y$-circle but are separated along 3-dimensional
$Z$-space. } \label{fig.1}
\end{figure}
\subsection{Simultaneous rolling and assisted inflation}
Our aim is to study the four dimensional cosmological solution
that one gets out of this system explained above while coupling to
four dimensional gravity. For this purpose, we will solve the
equations of motion derived from the effective action of the
`geometric tachyon' with the above potential for the D3-brane
falling in the background of NS5-brane on $R^3 \times S^1$ coupled
with four dimensional gravity action. The system is depicted in
the figure \eqn{fig.1}. In what follows we will only consider the
homogeneous mode. To make it look more familiar with that of the
open string tachyon effective action on unstable D-branes to the
quadratic order, we make a simple rescaling of the tachyon field
$T\rightarrow \sqrt{\alpha'}T$. After the rescaling the effective
action and the potential for the tachyon can be written as
\footnote{Though we are taking large number of D3-branes, the
NS5-branes are much heavier ($\sim \frac{1}{g^2_s}$) than the
Dp-branes ($\sim \frac{1}{g_s}$) in the weak string coupling
regime. For sufficiently weak coupling, the back reaction of these
probe branes can be ignored. We restrict ourselves to the
non-interacting geometric tachyon modes only in the DBI action.
Further, we ignore other excitations on the world volume of
D3-branes including gauge fields. Hence in the lowest order
analysis only geometric tachyons will contribute.} \bea S =
-\sum^{N}_{i =1} \int d^4 x V_i (T_i) \sqrt{-\det(g_{\mu\nu} +
\alpha' \p_{\mu} T_i  \p_{\nu} T_i)} \eea where in the action
above, all the potentials have same functional form, that is \bea
V_i(T_i) = V(T_i) = \tau_3 \left(1- {Q^2\over
{2\sqrt{\alpha'}T_i}}\right). \eea This form of potential is
however valid only in the asymptotic region where respective
$T_i$'s are very large. One can see that the equations of motion
for tachyon fields decoupled from each other. Hence we would
specifically like to assume that all D3-branes roll into the
NS5-brane at the same time. We will have the following
simultaneity ansatz for the purely time dependant configuration
\bea T_1 (t)&=& T_2 (t) = T_3(t) = \cdots = T_N (t) = \Phi(t) \cr
& \cr V_1(T_1)&=& V_2(T_2) = V_3(T_3) = \cdots = V_N(T_N) =
V(\Phi(t)) \label{ansatz1}\eea We shall be coupling this system to
that of the four dimensional gravity given by \bea S_{\rm{grav}} =
\frac{M^2_p}{2}\int{d^4 x \sqrt{-g} R}\ .\eea The field equations
we will solve combining gravity and the $N$ geometrical tachyon
effective action are as follows: \bea \ddot{\Phi}&=& -(1-\alpha'
\dot {\Phi^2})\left(M^2_s \frac{V_{,\Phi}}{V} + 3 H
\dot{\Phi}\right)\cr & \cr H^2 &=& \frac{8\pi G}{3}\left(1 -
\frac{k}{2\tilde{R}\Phi}\right)\frac{\tau_3 N}{\sqrt{1 - \alpha'
\dot{\Phi^2}}} \eea where $\tilde R = \frac{2 R'}{l_s}$. where we
have assumed the ansatze in \eqn{ansatz1}. As $k\ll 2\tilde{R}
\Phi$ and assuming $\alpha' \dot\Phi^2 \ll 1$, keeping only the
leading order terms we get \bea \ddot{\Phi} &=& - \left(M^3_s
\frac{Q^2}{2 \Phi^2} + 3 H \dot \Phi\right)  + O(\Phi^{-3}) \cr &
\cr H^2 &=& \frac{8\pi G \tau_3 N}{3}\left(1 - \frac{k}{2\tilde R
\Phi} + \frac{\alpha' {\dot \Phi}^2}{2}\right) \cr & \cr &=&
\frac{8 \pi G}{3} \frac{\tilde N}{g_s}\left( V_{\rm{eff}} +
\frac{\dot{\Phi}^2}{2}\right) ,\eea where in the last line we have
used \bea V_{\rm{eff}} (\Phi) = M^4_s \left(1 -
\frac{kM_s}{2\tilde R \Phi}\right), \>\>\> {\tilde N} =
\frac{N}{(2\pi)^3}, \>\>\> \tau_3 = \frac{1}{(2\pi)^3 g_s
{\alpha'}^2}, \>\>\> \Phi \rightarrow \sqrt{\alpha'} \Phi .\eea
One can also define a new field $\psi \equiv \sqrt{\frac{\tilde
N}{g_s}} \Phi $, which leads to the canonical form of equation
\bea H^2 = \frac{8 \pi G}{3}\left(V_{\rm{eff}} + \frac{{\dot
\psi}^2}{2}\right)\eea in which case
 \bea V_{\rm{eff}}(\psi) =
\frac{\tilde N M^4_s}{g_s}\left(1 - \sqrt{\frac{\tilde N}{g_s}}
\frac{k M_s}{2\tilde R \psi}\right), \>\>\> \eea This form of
potential will be used next to obtain various slow roll
quantities.

\subsection{Slow-roll parameters}
 The slow-roll parameters for this multi-field system are now
\cite{sasaki}
\bea
 \epsilon &=& {M_p^2 \over 2}\left({ V_{,\psi}\over V}\right)^2 \simeq
{  g_s\over 2\tilde  N  }({k\over 2\tilde R})^2 {M_p^2\over M_s^2}
\left( { M_s\over \Phi}\right)^4
\nonumber \\
\eta &=& {M_p^2 }{ V''\over V} \simeq -4
{ g_s\over 2\tilde  N }({2\tilde R\over k})^2 {M_p^2\over M_s^2}
 \left( {k M_s\over 2\tilde R \Phi }\right)^3 \ \eea
where $'$ indicate derivatives with respect to field $\psi$. The r.h.s are
reexpressed in terms of $\Phi$.
To little bit simplify these expressions, we now define a
new quantity$ {1\over \omega^2}=
{g_s\over \tilde  N }({ 2\tilde R\over k})^2 {M_p^2\over M_s^2}
 $. Then we can write \bea &&
\epsilon=-{1\over 4} \eta \left( {k M_s\over 2\tilde R
\Phi}\right),~~~~ \eta =-{2\over\omega^2} \left( {k M_s\over
2\tilde R \Phi}\right)^3 \label{slow2}\eea Let us note that we are
working in the asymptotic regime where \be\label{bound1} \left({k
M_s\over 2\tilde R \Phi}\right)\ll 1 \ee holds good. Note that
this can be achieved by adjusting $k,~R'$ and $\Phi$.\footnote{
The special advantage here compared to the case of
\cite{panigrahi}, where NS5-branes were on transverse $R^4$, is
that we could now vary the ratio ${k\over 2\tilde R}$ by varying
the radius $R'$  of $S^1$.} However, for the sake of simplifying
the analysis let us set from now onwards that during the slow roll
\be \left({k M_s\over 2\tilde R \Phi}\right)\simeq .01 .
\label{om1} \ee From the observational bounds on $\eta_{obs} \le
.02$, we determine that, following \eqn{om1} and \eqn{slow2}, we
should have
$$ \omega^2 \ge {2\over\eta_{obs}}\left({k M_s\over 2\tilde R
\Phi}\right)^3
\simeq (.01)^2 .$$
Thus to be reasonable we shall take $\omega^2\sim.001$ in the
following analysis.

Let us now look at the amplitudes of scalar perturbations. During
slow roll the square of
the amplitudes can be obtained as
\bea
\delta_s^2\simeq {1\over 150 \pi^2 \epsilon} {V_{eff}\over M_p^4}\approx
{1\over 150 \pi^2 \epsilon} \omega^2 ({2\tilde R\over k})^2
{M_s^2\over M_p^2}
\label{ampl1}\eea
Note that the amplitudes are bounded as $\delta_s^2
<10^{-10}$ \cite{wmap}.
Similarly, during slow roll the Hubble parameter is
\be
H^2\simeq {V_{eff} \over 3 M_p^2}\approx{\tilde N \over g_s} {M_s^4\over 3
M_p^2}=\omega^2 ({2\tilde R\over k})^2 {M_s^2\over 3}\ .
\ee
It is desirable to keep $H \ll M_s$, so that stringy effects are
suppressed.  Working  with $\omega^2 \approx .001$ (which is the safest to
choose
in case one chooses larger initial values of ${k M_s\over 2\tilde R
\Phi}$), the bound
$H\ll M_s$  can be achieved
so long as we ensure  ${k\over2\tilde R}\ge {1\over \sqrt{30}}\approx .18
$.
For this to happen, equation \eqn{om1} tells us that
$$\Phi\simeq 18 M_s\ .$$
It means that $D3$-branes have to be sufficiently far away from
the NS5-branes initially. This would correspond to the trans-stringy
situation.

Now, from eq. \eqn{om1} and the observational bounds on $\eta$ we
determine that
$$\epsilon \simeq {\eta\over4}{k\over 2\tilde R} {M_s\over\Phi}\sim
10^{-3} $$ and from \eqn{ampl1} we can find \bea \delta_s^2
\approx 10^{-2} {M_s^2\over M_p^2} \label{ampl2} \eea Since the
bounds on the amplitudes are  $\delta_s \le 1.9\times 10^{-5}$, it
fixes that the string scale to be in the range \be M_s \le
10^{-4}M_p \ .\ee Hence, the number of 3-branes goes as \be
{N\over (2\pi)^3}= g_s \omega^2 ({2\tilde R\over k})^2 {M_p^2\over
M_s^2} \sim g_s 10^6 \ .\ee
Actual number however will depend upon
the strength of weak string coupling in a given cosmological
vacuum. Nevertheless for a very weak string coupling the number of
3-branes required for assisted inflation could be very low indeed.
A value of  $g_s \sim 10^{-5}$  is
acceptable in our model, which makes the number of the D3-branes to be
small about $10^3$ or so. We note that   $g_s > 10^{-4}$ is not
acceptable in
our model.\footnote{ This is because in the probe approximation we must
have ${k \over g_s^2} \gg {N\over g_s} $, i.e. ${k\over N g_s} \gg1$.
Since $N\sim (2\pi)^3 g_s 10^6 \approx 2.5\times 10^8 g_s$, for $k=2$
we must have $g_s < 10^{-4}$ in order that probe approximation holds
good. Also a large number density of D3-branes may not be allowed
due to topological restrictions on CY coming from tadpole
charge cancellations.}

A few comments as compared to \cite{panigrahi}, where we considered
the D3-branes falling into the NS5-brane on transverse $R^4$, are
in order. First the number of probe D3-branes seem to be much
smaller (of the order $10^3$) than in the paper \cite{panigrahi}.
Further, in
the present case, we have the liberty of varying the ratio like
$\frac{k}{2\tilde R}$, by varying the radius of the circle
$S^1$, instead of varying $k$, that roughly fixes the radion
in terms of string mass $M_s$.

Before closing this section let us discuss few facts regarding the
4D physics out of
our system.
The typical relationship between 4D Planck mass and the string
mass is \bea M_p^2/M_s^2= v_0/((2\pi)^7 g_s^2) \label{f67}\eea
where $v_0= {\rm
Volume}/(l_s)^6 $ is the overall volume factor of the compact
3-fold. Presumably, to control stringy corrections this CY3
must have a large volume, $v_0\gg1$. We  consider the
asymmetric CY3 with the volume factor $v_0\sim 10^3$ which is
also preferred
by a small value of string coupling. Let us assume that the
directions along CY3 are $x^4, x^5, z_i, y$. Out of this we
make the assumption that the compact $y$-coordinate has radius
$R'>l_s$, and that size of $z$-coordinates goes as $|z| \gg l_s$
($|Z| \sim 18 l_s$).
We further note that for
$\frac{M_s}{M_p} \sim 10^{-4}$ and taking $g_s\sim 10^{-5}$ as determined
above,
we find from \eqn{f67} that $v_0 \sim  10^{13} g_s^2 \simeq 10^3$.
It also gives 3-brane density $N/v_0 <
1$. It is a quantity which says that there is no more than
one D3-brane per unit CY3 volume measured in the units of string
length. A higher density will induce backreaction on the CY3 in question.
With a volume cutoff, of course, we cannot take $Z$ to be very
large, because the displacement of 3-branes will further be
restricted by the size of the internal CY3. Further note that we
cannot take very small $g_s $ either, as that can reduce
the volume factor $v_0 \sim O(1)$ which is not preferred.
Hence $g_s \simeq 10^{-5}$ seems to be just
perfect choice.
We will show numerically in the next section that with the above
conditions, a
slow roll inflation is possible. However, there is the issue of
stabilization of various moduli that arises due to the
compactification. Strictly speaking all the moduli have to be
stabilized for a consistent inflationary model \cite{kklmmt}.
Studying the full fledged moduli stabilization with fluxes in this
particular
model is beyond the scope of the present paper. We wish to come
back to this issue in future.
In the next section we  study the effect of conserved angular
momentum in the system.

\section{Spin-flation}

We now consider a case where the rolling 3-branes have a spin
along the $R^3$ plane. The corresponding conserved angular
momentum will appear as scalar modes, $\theta^r$, in the DBI
action. The action will then read as
\begin{eqnarray}\label{fu1}
S_{3} &=&  -\tau_{3}\int
d^{3 + 1}~x \sqrt{-g} V(Z) \sqrt{1 + h (\p_a Z \p^a Z + \p_a
\theta^r \p^a \theta^s \tilde g_{rs}}).
\end{eqnarray}
where $V= \frac{1}{\sqrt{h(Z)}}$. The metric
$\tilde g_{rs}$ is the metric on the $\theta^r$ angular space.
In our notation, it is a
metric of the spherical polar coordinate type, $\tilde
g_{rs}={\rm diag}(Z^2,Z^2
\sin^2\theta_1)$.
We shall take for conserved angular
momenta, $l_r$, and the product as $l^2=l_r l_s \tilde g^{rs}$. The
conservation equation in the FRW background which follows from the above
action is
\be\label{fu2}
\gamma a(t)^3 h^{1/2} \dot\theta^r \tilde g_{rs}\equiv l_s
\ee
where $a(t)$ is the scale factor in the metric and $\gamma$ is the factor
generated by the non-standard kinetic term
\be\label{fu2a}
\gamma=\sqrt{(1+l^2/a^6)/(1-h \dot Z^2)}\ .
\ee
The gravitational equation is
\be\label{fu3}
{\dot a^2\over a^2}={8\pi G \over 3} \tau_3 V \gamma
\ee
this is also  the constraint equation. We also write down the
evolution equation
\be\label{fu4}
\dot H = -{4\pi G} \tau_3 {V\over\sqrt{1-h\dot Z^2}}{(h \dot Z^2 +
l^2/a^6) \over\sqrt{1+l^2/a^6}}
\ee
The $Z$-equation can be similarly obtained as
\bea\label{fu6}
\ddot Z + {h'\over 2 h}\dot Z^2+\dot{Z} (1-h\dot{Z}^2)\left(3 H + ({\dot
Z\over Z}+3H){l^2 a^{-6}\over (1+l^2 a^{-6})}\right) \br
 -{1\over 2 h \gamma^2} \left( {h'\over h} + {l^2 \over a^{6}}
( {h'\over h} + {2\over Z}(1-h\dot Z^2 )) \right)=0 \br
\eea

Considering now a simpler case when the D3-branes are moving in
an equatorial plane in $R^3$ transverse space. The branes are assumed to
be fixed along $S^1$. We will have
$\dot\theta_1\ne 0, ~\dot \theta_2=0$,
and let the orbital angular momentum in the
plane to be $L$. Then eq.\eqn{fu3} will reduce to
\be\label{fu5}
H^2={8\pi G \tau_3\over 3}  \sqrt{1+L^2 Z^{-2}a^{-6} \over h (1-h\dot
Z^2)}
\ee
As we notice that all $L$ dependence is contained in the $L^2$ term in the
above equation.
The $L$ dependent terms modify the tachyon rolling process by adding to
the initial value of Hubble parameter $H$. We notice that when $L=0$ these
equations directly reduce to  zero angular momentum case.
   The equations \eqn{fu4} to  \eqn{fu6} shall be used for studying
inflationary numerical solutions.

\subsection{Numerics}
We shall be considering the case of orbital motion in the
$(Z,\theta)$ plane as discussed in the previous section. Let us
take the orbital momentum to be $L$. For studying the numerical
aspects we the use following equations of motion \bea\label{nfu1}
&&\ddot Z + {h'\over 2 h}\dot Z^2+\dot{Z} (1-h\dot{Z}^2)\left(3 H
+ ({\dot Z\over Z}+3H){L^2 Z^{-2}a^{-6}\over (1+L^2
Z^{-2}a^{-6})}\right) \br && -{1\over 2 h \gamma^2} \left(
{h'\over h} + {L^2 \over Z^{2}a^{6}} ( {h'\over h} + {2\over
Z}(1-h\dot Z^2 )) \right)=0 ~,\br &&\dot a(t)^2={q\over 3} a(t)^2
\sqrt{1+L^2 Z^{-2}a^{-6} \over h (1-h\dot Z^2)} \eea where we
defined $q\equiv{8\pi G N \tau_3}$ and we shall be using
$h(Z)=1+{k l_s\over 2\tilde{R} Z}$. Considering the fact that the
BPS D3-branes are initially very far away, $Z(0)\gg R'> l_s$, and
roll down with zero initial velocity, $\dot Z(0)\simeq 0$, the
initial value of Hubble parameter is given by ${q\over 3}
\sqrt{1+L^2 Z(0)^{-2}a(0)^{-6}}$. We shall be setting $a(0)=1$ and
it is useful to have ${L \over Z(0)}$  finite in order to study
the effect of orbital motion on the inflationary dynamics. As the
universe will have an accelerated expansion the $L$ dependant
terms will be heavily suppressed during evolution even though $Z$
will decrease in time as we shall see. Thus the essential
contribution from $L$ terms will come at the time of initial epoch
only.
\begin{figure}[!ht]
\leavevmode
\begin{center}
\epsfysize=6cm
\epsfbox{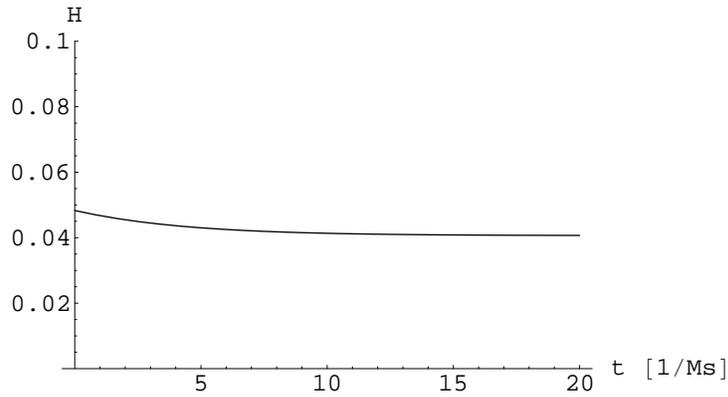}
\end{center}
\caption{\it $H(t)$ as a function of time for initial epoch of time. It is
clear that initial higher value that $L$ terms dominate the
expansion and later on $H$
sets to its plateau value during inflation. }
\label{fig.2}
\end{figure}
\begin{figure}[!ht]
\leavevmode
\begin{center}
\epsfysize=6cm
\epsfbox{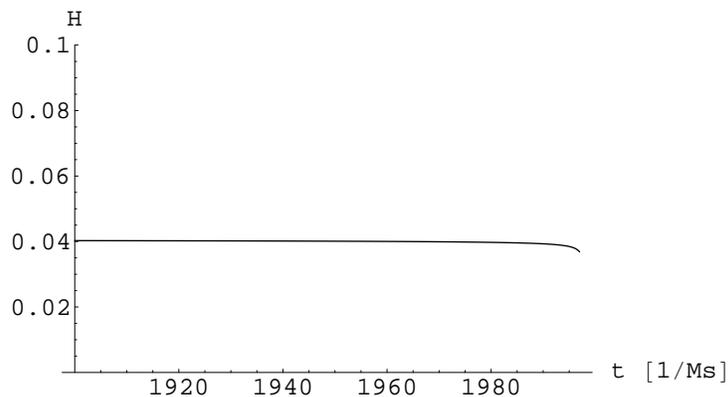}
\end{center}
\caption{\it Inflation ends at about the time when $t\sim 2000
M_s^{-1}$. From the area under the $H$-curve we estimate the
number of e-folds being around 80.} \label{fig.3}
\end{figure}
\begin{figure}[!ht]
\leavevmode
\begin{center}
\epsfysize=6cm
\epsfbox{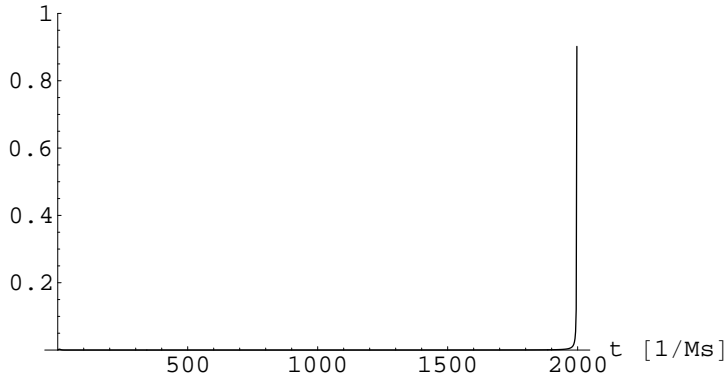}
\end{center}
\caption{\it The plot of  $h(Z)\dot Z^2 \equiv \dot T^2$. It shows
that tachyon velocity $\dot T^2$ tends to 1 when inflation ends. It
is the time when tachyon  condensation takes place. }
\label{fig.4}
\end{figure}

Keeping with our discussion on the slow-roll in the previous sections, we
can take
$$q={1\over 200},~
{k\over\tilde R}=0.8,~ L=20\ .$$ \footnote{ The value ${k \over
\tilde R}=.8$ confirms to a situation where we can have around 2
NS5-branes spread over the transverse circle of radius $R'=1.25
l_s$ or 4 NS5-branes spread over the transverse circle of radius
$R'=2.5 l_s$ .} We shall work in the units of $\alpha'=l_s^2=1$.
The initial values are taken as $\dot Z(0)=-.003,~ Z(0)=20,~
a(0)=1$ so that the bounds $Z\gg R'>l_s$ are respected. The
evolutions of Hubble parameter and velocities are plotted in
figures \eqn{fig.2} to \eqn{fig.4} for various intervals of time.
The number of e-folds could be estimated by finding out total area
under the $H(t)$ curve in the plateau regions. We find that the
number of e-folds are more when $L\ne 0$, but this number does not
change very much even if we take $L=20$. For a comparison, in the
figures \eqn{fig.6} to \eqn{fig.9}, we have  plotted $H(t)$   for
$L=0$ case also while keeping everything else the same. The
inflation ends  slightly earlier in the $L=0$ case as compared to
the $L=20$ case. This behaviour is on the expected lines since the
D3-branes spin around the NS5-brane as they slide towards them.
But due to accelerated expansion the effect of $L/a^3$ terms dies
out much faster. The visible effect of the angular momentum
appears only near to
 the beginning of the rolling process.
In the absence of any angular momentum the branes will
slide faster thus ending the inflation a little
early. We also find that the number of e-folds increases marginally if we
choose  smaller initial value of $\dot Z(0)$ which is expected.
Note that we have
chosen the values of various parameters such that  $H$ remains  throughout
very small compared to
string mass. This is along the lines of discussion in the
previous section so that stringy effects are suppressed.
\begin{figure}[!ht]
\leavevmode
\begin{center}
\epsfysize=6cm
\epsfbox{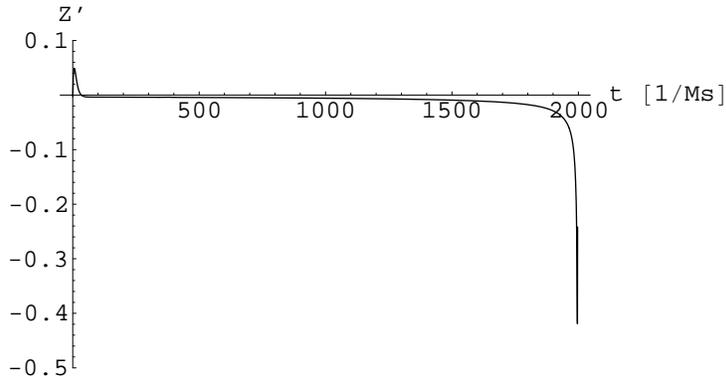}
\end{center}
\caption{\it The negative radion velocity $\dot Z$ indicating that
3-branes are
moving towards NS5-branes. The positive initial values indicate that due
to angular-momentum effect the D-branes initially move away from
NS-branes. When effect of
$L/a^3$ terms is diluted away due to expansion of space the D-branes again
fall towards the center.}
\label{fig.5} \end{figure}
\begin{figure}[!ht]
\leavevmode
\begin{center}
\epsfysize=6cm
\epsfbox{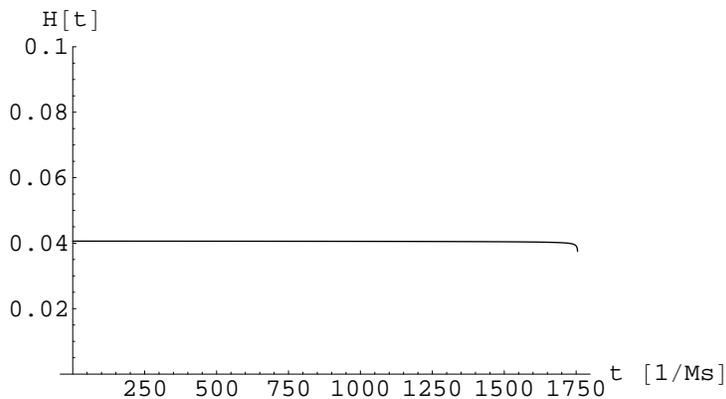}
\end{center}
\caption{\it $H(t)$ graph for $L=0$ case. The inflation
ends slightly early in this case at about $t=1750/M_s$. The number of
e-folds is reduced to 70.}
\label{fig.6}
\end{figure}
\begin{figure}[!ht]
\leavevmode
\begin{center}
\epsfysize=6cm
\epsfbox{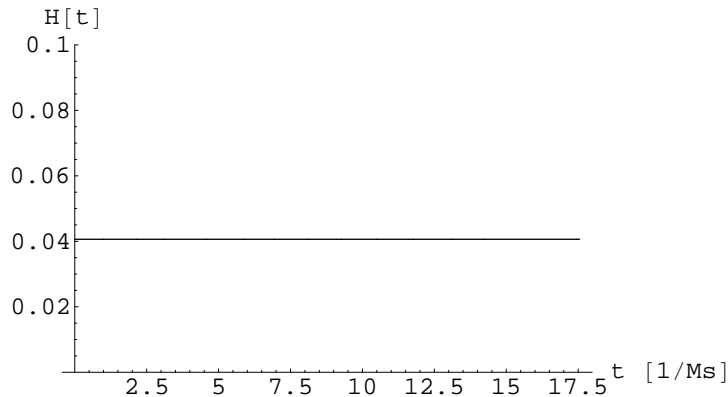}
\end{center}
\caption{\it $H(t)$  for initial epoch of time when
$L=0$. We compare it with $L=20$ case.}
\label{fig.7}
\end{figure}
\newpage
\begin{figure}[!ht]
\leavevmode
\begin{center}
\epsfysize=6cm
\epsfbox{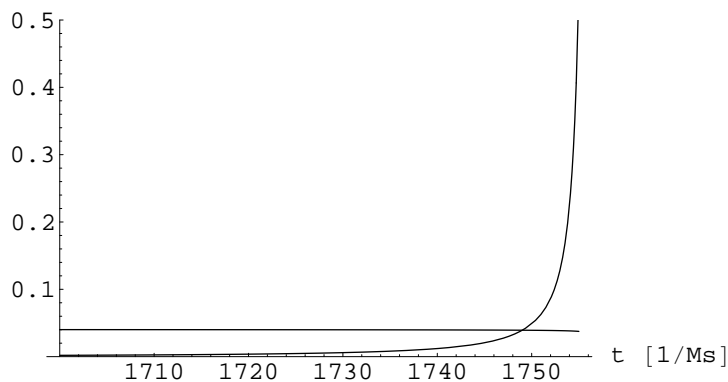}
\end{center}
\caption{\it $H$ and $h(Z)\dot Z^2$ (the rising curve)  for $L=0$. }
\label{fig.8}
\end{figure}
\begin{figure}[!ht]
\leavevmode
\begin{center}
\epsfysize=6cm \epsfbox{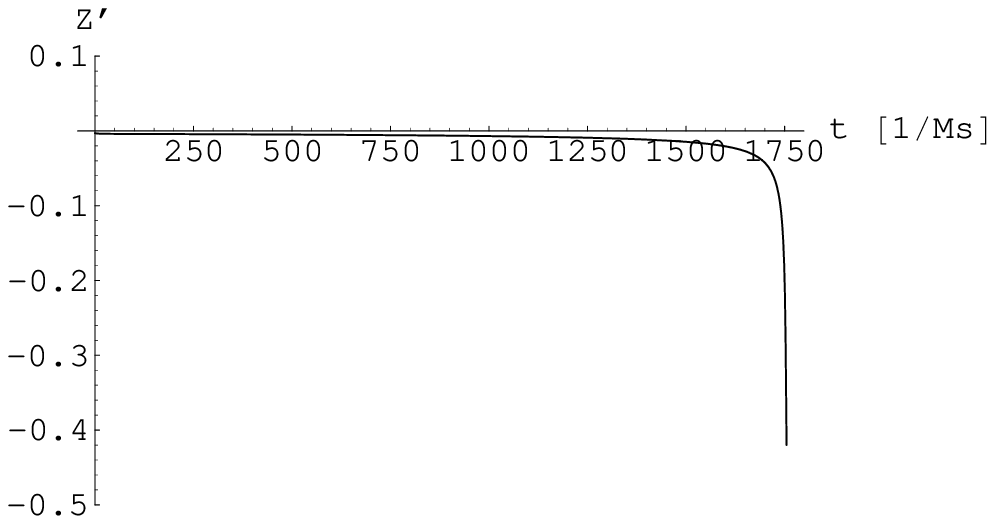}
\end{center}
\caption{\it  $\dot Z$ as a  for $L=0$. The D-branes smoothly fall
towards the center.}
\label{fig.9}
\end{figure}
\sect{Conclusions} We have studied, in this paper, the assisted
inflation from the homogeneous rolling of the D3-branes into the
NS5-branes on $R^3\times S^1$ coupled to four dimensional gravity,
where the branes are distributed along the $S^1$. We have shown
that the number of D3-branes needed for the assisted inflation are
indeed very low, for a sufficiently weak string coupling and for a
small radius of the $S^1$ circle. The number of e-foldings and
other slow roll parameters are shown to be similar to that of
N-tachyon assisted inflation. We have further studied the assisted
inflation when there is a conserved angular momentum for
D3-branes along a plane in  $R^3$. We have studied this effect
of spin numerically. We have shown that the effect of angular momentum
indeed slows down the process of rolling, which is not very surprising, as
the D-branes while slide into the NS5-branes, also spin along a
plane. The inflation has been shown to end at a later time
compared to the zero angular momentum case. This effectively adds more
e-folds to the inflation.  However, the effect of spin
can be observed for a very short duration at the initial stage of
inflation only. So this may not have any effect on the last 50-60
e-folds of inflation although overall number of e-folds will increase.

\par This results of the present paper can be extended in few ways. A
simple exercise could be to extend this idea of spin-flation in
\cite{panigrahi} to add conserved angular momentum and see how
exactly it affects the inflation. It would also be interesting to
include spin in the brane inflation in other warped backgrounds
and study the process.

\end{document}